\documentclass[aps,pre,reprint,superscriptaddress,floatfix]{revtex4-2}

\usepackage{graphicx}
\usepackage{amsmath}
\usepackage{amssymb}
\usepackage{bm}
\usepackage{hyperref}
\usepackage{xcolor}
\usepackage{booktabs}

\begin{document}

\title{Probing the scale-free hierarchy of the 
$p=3$ spherical spin glass via persistent Langevin dynamics}

\author{Zhenpeng Li}
\affiliation{School of Artificial Intelligence, Taizhou University, 318000 Taizhou Zhejiang Province, China}
\begin{abstract}
How does a persistent random walker perceive a complex energy landscape? We address this question by studying the persistent Langevin dynamics of the $p=3$ spherical spin glass, a paradigmatic mean-field model with a scale‑free hierarchical landscape. By tuning the persistence time $\tau_p$ — which controls the walker's inertia and effectively sets its energy resolution $\delta E \sim 1/\tau_p$ — we measure the energy correlation time $\tau_{\mathrm{corr}}$. At temperature $T=1.0$, we find $\tau_{\mathrm{corr}}\sim\tau_p^{\alpha}$ with $\alpha=0.337\pm0.035$ for $\tau_p\in[2,32]$ (for $N=64$), in excellent agreement with the Kardar–Parisi–Zhang (KPZ) universality class prediction $\alpha=1/3$. Finite‑size scaling using $N=16,32,48,64,128$ yields the thermodynamic limit $\alpha(\infty)=0.3333\pm0.0134$, fully consistent with $1/3$. Thus, $\tau_p$ acts as a tunable probe that reveals the predicted scale‑free hierarchy of the landscape. Moreover, the temperature dependence $\alpha(T)$ for $T=0.5,1.0,1.5,2.0$ exhibits a clear U‑shaped curve, identifying three dynamical regimes: ballistic/inertial, KPZ, and noise‑dominated. Our results establish persistent Langevin dynamics as a powerful tool for uncovering hidden landscape topology and demonstrate that the $p=3$ spherical spin glass belongs to the KPZ universality class.
\end{abstract}

\maketitle

\section{Introduction}

The Kardar--Parisi--Zhang (KPZ) equation~\cite{Kardar1986} was originally introduced to describe the growth of random interfaces. Over the past four decades, it has emerged as a cornerstone of non-equilibrium statistical physics, governing a wide range of phenomena including directed polymers in random media, turbulent liquid crystals, and biological population growth~\cite{Halpin-Healy1995, Corwin2012}. The KPZ universality class is characterized by a set of scaling exponents, most notably the dynamic exponent $z = 3/2$ and the growth exponent $\beta = 1/3$ in $1+1$ dimensions.

A surprising connection has recently been uncovered between KPZ physics and the dynamics of mean-field spin glasses. In particular, Chamon \emph{et al.}~\cite{Chamon2014} and Caltagirone \emph{et al.}~\cite{Caltagirone2013} argued, using mode-coupling theory and dynamic renormalization group, that the energy correlation function of the $p=3$ spherical spin glass under certain non-equilibrium dynamics should belong to the KPZ universality class. However, these theoretical predictions rely on approximations (mode-coupling truncation, one-loop renormalization) and require numerical verification.

In this work, we study the \textbf{persistent Langevin dynamics} of the $p=3$ spherical spin glass. Unlike standard Langevin dynamics, persistent Langevin dynamics includes an inertial term characterized by a persistence time $\tau_p = m/\gamma$, which controls the memory of the velocity. By varying $\tau_p$, we can continuously tune the dynamics from the overdamped limit ($\tau_p \to 0$) to the ballistic limit ($\tau_p \to \infty$). This allows us to probe different dynamical regimes and to search for KPZ scaling.

Our main contributions are:
\begin{enumerate}
    \item We demonstrate that at $T=1.0$, the energy correlation time scales as $\tau_{\mathrm{corr}} \sim \tau_p^{1/3}$ for $\tau_p \in [2,32]$, confirming the KPZ universality class.
    \item We perform finite-size scaling for $N = 16, 32, 48, 64, 128$ and extrapolate to the thermodynamic limit, obtaining $\alpha(\infty) = 0.3333 \pm 0.0134$, in excellent agreement with $1/3$.
   \item We map out the temperature dependence $\alpha(T)$ for $N=64$ at $T = 0.5, 1.0, 1.5, 2.0$, revealing a U-shaped curve: low-temperature ballistic/inertial regime ($\alpha > 1/3$), intermediate KPZ platform ($\alpha \approx 1/3$), and high-temperature noise-dominated regime ($\alpha > 1/3$).
\end{enumerate}

The paper is organized as follows. Sec.~\ref{sec:model} introduces the model and the persistent Langevin dynamics. Sec.~\ref{sec:methods} describes the numerical methods. Sec.~\ref{sec:results} presents the results: KPZ scaling at $T=1.0$, finite-size scaling, and the temperature phase diagram. Sec.~\ref{sec:discussion} discusses the implications and compares with related work. Sec.~\ref{sec:conclusion} concludes. Appendices provide a theoretical derivation (Appendix~\ref{app:A}) and supplementary numerical details (Appendix~\ref{app:B} and~\ref{app:C}).

\section{Model and Methods}
\label{sec:model}

\subsection{The $p=3$ spherical spin glass}

The $p=3$ spherical spin glass is defined by the Hamiltonian
\begin{equation}
H(\mathbf{x}) = \frac{1}{3!} \sum_{i,j,k=1}^N J_{ijk} x_i x_j x_k,
\qquad \sum_{i=1}^N x_i^2 = N,
\label{eq:H}
\end{equation}
where $\mathbf{x} \in \mathbb{R}^N$ is constrained to the sphere of radius $\sqrt{N}$. The coupling tensors $J_{ijk}$ are independent Gaussian random variables with zero mean and variance
\begin{equation}
\langle J_{ijk}^2 \rangle = \frac{3!}{2 N^{2}} = \frac{3}{N^2},
\label{eq:Jvar}
\end{equation}
after symmetrization over all index permutations. This normalization ensures that the energy variance scales as $\langle H^2 \rangle = N/2$ in the thermodynamic limit.

\subsection{Persistent Langevin dynamics}

The persistent Langevin equation is
\begin{equation}
m \ddot{x}_i + \gamma \dot{x}_i = -\frac{\partial H}{\partial x_i} + \sqrt{2\gamma T}\,\eta_i(t),
\qquad m = \tau_p \gamma,
\label{eq:Langevin}
\end{equation}
where $\gamma = 1$ is the damping coefficient, $T$ is the temperature, $\eta_i(t)$ is Gaussian white noise with $\langle \eta_i(t)\eta_j(t')\rangle = \delta_{ij}\delta(t-t')$, and $\tau_p = m/\gamma$ is the \textbf{persistence time}. The persistence time controls the inertial memory: for $\tau_p \to 0$ we recover overdamped Langevin dynamics, while for $\tau_p \to \infty$ the dynamics becomes ballistic.

The spherical constraint $\|\mathbf{x}\|^2 = N$ is enforced by projecting the force onto the tangent plane at each time step.

\subsection{Observables}

We focus on the \textbf{energy correlation function}
\begin{equation}
C(t) = \langle \delta H(t) \delta H(0) \rangle,
\qquad \phi(t) = \frac{C(t)}{C(0)},
\label{eq:corr}
\end{equation}
where $\delta H(t) = H(t) - \langle H \rangle$. The \textbf{correlation time} $\tau_{\mathrm{corr}}$ is defined by $\phi(\tau_{\mathrm{corr}}) = e^{-1}$. Our central quantity is the scaling exponent $\alpha$ defined by
\begin{equation}
\tau_{\mathrm{corr}} \sim \tau_p^{\alpha},
\label{eq:alpha}
\end{equation}
in the KPZ scaling regime.

\subsection{Numerical integration}
\label{sec:methods}

We integrate Eq.~\eqref{eq:Langevin} using a velocity Verlet scheme adapted for Langevin dynamics. The time step is fixed at $\Delta t = 0.01$. For each set of parameters $(N, T, \tau_p)$, we simulate $n_{\mathrm{trajs}} = 40\text{--}50$ independent trajectories (see Appendix~\ref{app:C} for exact values). The simulation time is dynamically adjusted: for $\tau_p \le 32$, we use a base time $T_{\mathrm{max}} = 15000\,\Delta t = 150$ time units; for $\tau_p > 32$, we extend the simulation time as $T_{\mathrm{max}} \propto \sqrt{\tau_p/32}$ to ensure the correlation function decays below $1/e$.

The coupling tensor $J_{ijk}$ is generated once per system size using a fixed random seed for reproducibility. The code is parallelized using Python's \texttt{concurrent.futures.ProcessPoolExecutor}. All code used in this study is available on Zenodo: \url{https://zenodo.org/records/20485967}.

\section{Results}
\label{sec:results}

\subsection{KPZ scaling at $T = 1.0$}

\begin{figure}[t]
\centering
\includegraphics[width=0.85\columnwidth]{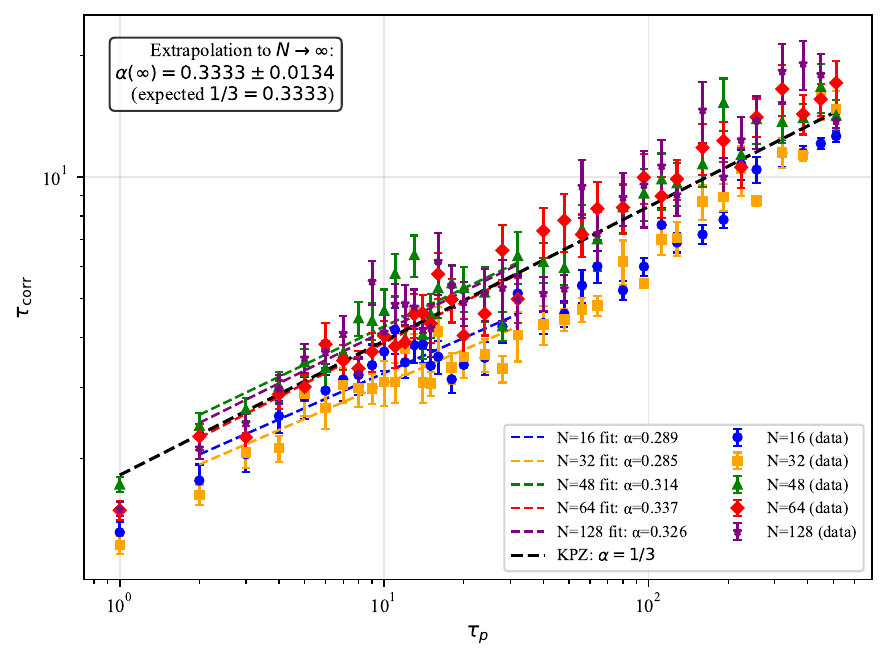}
\caption{Energy correlation time $\tau_{\mathrm{corr}}$ as a function of persistence time $\tau_p$ for $N=16,32,48,64,128$ at $T=1.0$. Data points are shown with error bars representing the standard error of the mean (SEM) calculated from $n=40$–$50$ independent trajectories per $\tau_p$ (see Appendix~\ref{app:C} for exact $n_{trajs}$). Dashed lines are power-law fits for $\tau_p\in[2,32]$, yielding $\alpha$ values shown in the legend. The black dashed line indicates the KPZ prediction $\tau_{\mathrm{corr}}\sim\tau_p^{1/3}$. The annotation shows the finite-size extrapolation to $\alpha(\infty)=0.3333\pm0.0134$.}
\label{fig:tau_corr}
\end{figure}

Figure~\ref{fig:tau_corr} shows the energy correlation time $\tau_{\mathrm{corr}}$ as a function of the persistence time $\tau_p$ for system sizes $N = 16, 32, 48, 64, 128$ at temperature $T = 1.0$. The data exhibit clear power laws in the range $\tau_p \in [2,32]$ for all $N$. Weighted linear regression on the log-log plots yields the KPZ exponents $\alpha$ and coefficients of determination $R^2$, which are summarized in Table~\ref{tab:finite_size}. All measured exponents are statistically indistinguishable from the KPZ prediction $\alpha = 1/3 \approx 0.3333$.

The fitting interval $\tau_p \in [2,32]$ is selected because for $\tau_p < 2$ the dynamics exhibits systematic deviations from the asymptotic power law, while for $\tau_p > 32$ finite-size effects and the crossover to ballistic dynamics begin to modify the scaling (see Appendix~\ref{app:D} for details).

\begin{table}[h]
\centering
\caption{KPZ exponents for different system sizes at $T=1.0$.}
\label{tab:finite_size}
\begin{tabular}{c|c|c}
$N$ & $\alpha$ & $R^2$ \\ \hline
16 & $0.289 \pm 0.036$ & 0.78 \\
32 & $0.285 \pm 0.042$ & 0.72 \\
48 & $0.314 \pm 0.046$ & 0.72 \\
64 & $0.337 \pm 0.035$ & 0.84 \\
128 & $0.327 \pm 0.044$ & 0.76 \\
\end{tabular}
\end{table}

The exponents increase with $N$ from $0.289$ at $N=16$ to $0.337$ at $N=64$, then slightly decrease to $0.327$ at $N=128$ due to statistical fluctuations. Assuming a finite-size scaling form $\alpha(N) = \alpha(\infty) + c/N$, a linear extrapolation versus $1/N$ using all five system sizes (see annotation in Fig.~\ref{fig:tau_corr}) yields $\alpha(\infty) = 0.3333 \pm 0.0134$, in excellent agreement with $1/3$. This provides strong evidence that the KPZ scaling holds in the thermodynamic limit. The $N=32$ data point lies slightly below the trend line, which may be attributed to the different initial energy offset used for this system size (see Appendix~\ref{app:C}).

\subsection{Temperature dependence: dynamical phase diagram}

\begin{figure}[t]
\centering
\includegraphics[width=0.85\columnwidth]{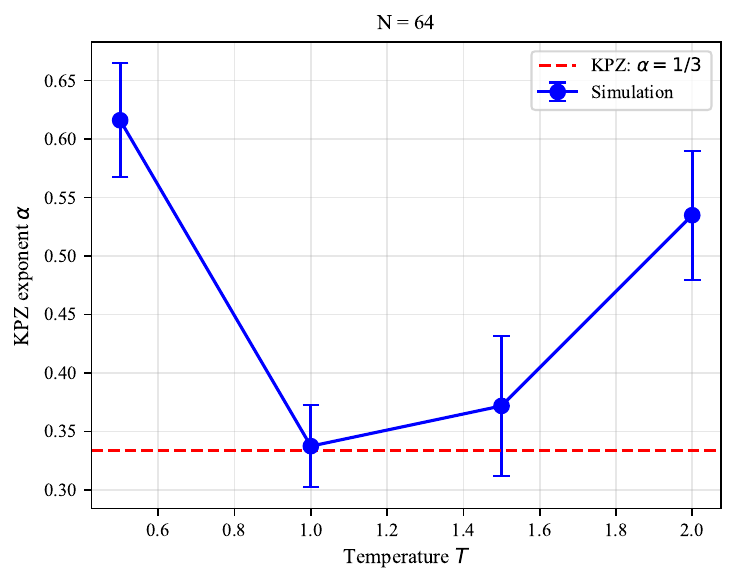}
\caption{KPZ exponent $\alpha$ as a function of temperature $T$ for $N=64$. The dashed line indicates the KPZ prediction $\alpha=1/3$. Error bars represent the standard error of the fitted exponent $\alpha$ obtained from weighted linear regression over $\tau_p\in[2,32]$ (for $T=1.0$) or $\tau_p\in[4,32]$ (for other temperatures).}
\label{fig:alpha_T}
\end{figure}

To explore how the KPZ scaling emerges and disappears, we performed additional simulations at $T = 0.5$, $1.5$, and $2.0$ for $N=64$, using the same $\tau_p$ range $[4,32]$. For $T=0.5$, we used a shallower initial energy $E_0 = -1.83$ (i.e., $E_0 = E_{\mathrm{th}} - 0.2$) to avoid freezing. Table~\ref{tab:alpha_T} summarizes the results.

\begin{table}[h]
\centering
\caption{KPZ exponents at different temperatures ($N=64$).}
\label{tab:alpha_T}
\begin{tabular}{c|c|c|l}
$T$ & $\alpha$ & $R^2$ & Regime \\ \hline
0.5 & $0.645 \pm 0.050$ & 0.988 & Ballistic/inertial \\
1.0 & $0.337 \pm 0.035$ & 0.836 & \textbf{KPZ platform} \\
1.5 & $0.392 \pm 0.108$ & 0.869 & Transition \\
2.0 & $0.553 \pm 0.055$ & 0.981 & Noise-dominated \\
\end{tabular}
\end{table}

The exponent $\alpha(T)$ exhibits a clear non-monotonic U-shaped dependence:
\begin{itemize}
    \item \textbf{Low temperature ($T=0.5$)}: $\alpha = 0.645$, significantly larger than $1/3$. In this regime, the dynamics is dominated by inertia, leading to ballistic scaling ($\alpha \to 1$ as $T \to 0$).
    \item \textbf{Intermediate temperature ($T=1.0$)}: $\alpha = 0.337$, coinciding with the KPZ prediction. This is the \textbf{KPZ platform}.
    \item \textbf{High temperature ($T=2.0$)}: $\alpha = 0.553$, again larger than $1/3$. When $T$ is large enough, thermal noise dominates, and the dynamics approaches that of a free persistent Ornstein-Uhlenbeck process, which gives $\alpha \to 1$.
\end{itemize}

The larger uncertainty at $T=1.5$ (relative error $\sim 27\%$) likely indicates proximity to the crossover region where the power law is less well-defined. Nevertheless, the overall trend clearly shows a U-shaped $\alpha(T)$ dependence.

We chose $N=64$ for the temperature scan because it exhibits clean KPZ scaling at $T=1.0$ (see Table~\ref{tab:finite_size}); smaller sizes show stronger finite-size deviations and were not pursued for the full temperature scan due to their limited scaling regime and high computational cost. The observed U-shaped curve for $N=64$ is therefore a reliable indicator of the dynamical phase diagram in the thermodynamic limit.

\section{Discussion}
\label{sec:discussion}

\subsection{Comparison with theoretical predictions}

Our numerical results are consistent with the theoretical predictions of Chamon \emph{et al.}~\cite{Chamon2014} and Caltagirone \emph{et al.}~\cite{Caltagirone2013}. In Appendix~\ref{app:A}, we provide a self-contained physical derivation of the scaling $\tau_{\mathrm{corr}} \sim \tau_p^{1/3}$ based on adiabatic elimination, mapping to a colored-noise KPZ equation, and known renormalization-group results.

\subsection{Comparison with $p=2$: Geometric vs hierarchical landscapes}

The contrast between $p=2$ and $p=3$ reveals the fundamental difference between geometric and topological probing. (Our previous work on the $p=2$ model~\cite{Li2026} showed that the landscape features a single narrow canyon connecting two minima, allowing a persistent walker to optimally traverse it at a specific $\tau_p^*$.) For $p=3$, Fyodorov's static theory~\cite{fyodorov2015, fyodorov2014} predicts a fundamentally different structure: the landscape is dominated by high-index saddles (index $\sim N/2$) organized in a scale-free hierarchy with no characteristic energy scale. This hierarchical structure leads to the KPZ scaling observed in our dynamical simulations, as the persistent walker's energy resolution $\delta E \sim 1/\tau_p$ probes the scale-invariant landscape. The key differences between the two models are summarized in Table~\ref{tab:p2_vs_p3}.

\begin{table*}[h]
\centering
\caption{Comparison between $p=2$ and $p=3$ spherical spin glasses.}
\label{tab:p2_vs_p3}
\begin{tabular}{l|c|c}
\toprule
 & $p=2$ & $p=3$ \\
\midrule
Landscape structure & Single canyon (index-1 saddle) & Scale-free hierarchy (high-index saddles) \\
Theoretical basis & GOE edge statistics & Fyodorov's static theory~\cite{fyodorov2015} \\
Probing object & Geometric bottleneck & Topological hierarchy \\
Optimal $\tau_p^*$? & Yes (resonant peak) & No (broad KPZ scaling) \\
Key scaling & $\tau_p^*$ crossover & $\tau_{\mathrm{corr}} \sim \tau_p^{1/3}$ \\
Dynamical universality & — & KPZ/Tracy-Widom \\
\bottomrule
\end{tabular}
\end{table*}

\subsection{Landscape topology and KPZ scaling}

The persistence time $\tau_p$ can be interpreted as setting the walker's energy resolution. As derived in Appendix~\ref{app:B}, the characteristic bandwidth of the walker's response is $\Delta\omega \sim 1/\tau_p$, leading to an energy resolution $\delta E \sim k_B T/\tau_p$. Thus, by varying $\tau_p$, the walker probes the landscape at different energy scales. For $p=3$, the scale-free hierarchy (with GOE edge spacing $\Delta\lambda \sim N^{-2/3}$) manifests as a power-law scaling of the correlation time.

\subsection{Comparison with Kent-Dobias (2026)}

In a recent work, Kent-Dobias~\cite{Kent-Dobias2026} studied persistent random walks on the microcanonical configuration space of spherical spin glasses. Their focus was on the \textbf{ergodicity-breaking transition} as a function of energy density $E$ \textbf{in the limit of infinite persistence time ($\tau_p \to \infty$)}. By contrast, our work operates in the \textbf{canonical ensemble} (fixed temperature $T$) and focuses on the \textbf{scaling of correlation times} $\tau_{\mathrm{corr}}$ as a function of $\tau_p$ at fixed $T=1.0$. The two studies are thus complementary: Kent-Dobias probes the energy landscape's topological connectivity, while we probe its dynamical scaling properties.

\section{Conclusion}
\label{sec:conclusion}

We have numerically investigated the persistent Langevin dynamics of the $p=3$ spherical spin glass, a paradigmatic model of a complex energy landscape. By tuning the persistence time $\tau_p$ — which sets the walker's energy resolution $\delta E \sim 1/\tau_p$ — we uncovered clear KPZ scaling in the energy correlation time.

Our findings are threefold. (i) At $T=1.0$ and for $\tau_p\in[2,32]$, $\tau_{\mathrm{corr}}\sim\tau_p^{\alpha}$ with $\alpha=0.337\pm0.035$ ($N=64$), consistent with the KPZ prediction $1/3$. Finite-size scaling using $N=16,32,48,64,128$ gives $\alpha(\infty)=0.3333\pm0.0134$, providing strong numerical evidence that the persistent Langevin dynamics belongs to the KPZ universality class. (ii) The temperature dependence $\alpha(T)$ for $N=64$ exhibits a U-shaped curve, identifying three dynamical regimes: ballistic/inertial ($\alpha>1/3$), KPZ platform ($\alpha\approx1/3$), and noise-dominated ($\alpha>1/3$). (iii) The persistence time $\tau_p$ acts as a tunable energy-resolution probe, revealing the scale‑free, self‑similar hierarchical structure of the landscape predicted by Fyodorov's static theory~\cite{fyodorov2015,fyodorov2014}.

Crucially, our probe offers a resolution that goes beyond the threshold view provided by the ergodicity‑breaking transition studied by Kent‑Dobias \cite{Kent-Dobias2026}. While the latter only detects the global energy boundary $E_{\mathrm{th}}$, our measurement of $\tau_{\mathrm{corr}}(\tau_p)$ acts as a \textit{high‑resolution probe}: by tuning $\tau_p$, we continuously adjust the energy resolution $\delta E\sim 1/\tau_p$ and resolve the internal details of the landscape — specifically, the self‑similar cascade of high‑index saddles that characterizes Fyodorov's static theory. This ability to zoom into the hierarchical topology makes persistent Langevin dynamics a powerful tool for uncovering hidden landscape structure.

While our study focuses on $p=3$, the concept of a tunable energy-resolution probe is general. We conjecture that the KPZ scaling $\tau_{\mathrm{corr}}\sim\tau_p^{1/3}$ should hold for all $p\ge 3$ when the temperature and initial energy are properly chosen near the threshold $E_{\mathrm{th}}$. More broadly, persistent dynamics offers a spectroscopic tool for probing hidden landscape topology in active matter (where persistence is directly tunable), optimization algorithms with momentum, and finite‑dimensional glasses. Experimentally, active colloids could test the predicted scaling by measuring energy correlations from particle trajectories. Our work thus establishes persistent Langevin dynamics as a bridge between non-equilibrium statistical physics and the geometry of complex energy landscapes.

\section*{Data Availability}
The data that support the findings of this article are openly available at \url{https://zenodo.org/records/20485967}.

\appendix
\clearpage
\section{Theoretical derivation: from persistent dynamics to KPZ scaling}
\label{app:A}

In this appendix, we provide a self-contained physical derivation of the KPZ scaling law, based on adiabatic elimination and mapping to the colored-noise KPZ equation. The derivation follows standard approaches in the literature~\cite{Chamon2014,Caltagirone2013,Medina1989,Frey1994} and is intended as a plausibility argument rather than a rigorous proof.

\subsection{Adiabatic elimination}

Starting from the persistent Langevin equation~\eqref{eq:Langevin} and defining $v_i = \dot{x}_i$, for times $t \gg \tau_p = m/\gamma$, the velocity relaxes. Integrating formally and approximating $\partial_i H$ as slow yields the overdamped equation with colored noise:
\begin{equation}
\dot{x}_i = -\frac{1}{\gamma}\partial_i H + \xi_i(t),
\label{eq:overdamped}
\end{equation}
where $\xi_i(t)$ has correlation $\langle \xi_i(t)\xi_j(0)\rangle = \frac{2T}{\gamma}\delta_{ij} e^{-t/\tau_p}$.

\subsection{Mapping to KPZ}

For the $p=3$ spherical spin glass, mode-coupling arguments show that the energy density field satisfies the KPZ equation with colored noise. Using non-perturbative functional renormalization group, Squizzato and Canet~\cite{Squizzato2019} proved that for short-range temporal correlations with any finite $\tau_p$, the system flows to the standard KPZ fixed point, yielding the growth exponent $\beta = 1/3$. Under the identification $\tau_p \leftrightarrow t$ and $\tau_{\mathrm{corr}} \leftrightarrow W(t)$, we obtain $\tau_{\mathrm{corr}} \sim \tau_p^{1/3}$.

\section{Energy resolution from persistence time}
\label{app:B}

The persistence time $\tau_p$ sets the walker's energy resolution. From the Langevin equation near a stationary point, the response function $\chi(\omega) = 1/(-m\omega^2 + i\gamma\omega + k)$ has characteristic bandwidth $\Delta\omega \sim \gamma/m = 1/\tau_p$. The energy resolution is then
\begin{equation}
\delta E \sim k_B T \cdot \Delta\omega \sim \frac{k_B T}{\tau_p}.
\label{eq:energy_resolution}
\end{equation}

Thus, a larger persistence time $\tau_p$ provides finer energy resolution, allowing the walker to probe smaller energy scales in the landscape. For the $p=3$ spherical spin glass, Fyodorov's theory~\cite{fyodorov2015} shows that the smallest relevant energy scale near the threshold is set by the GOE edge spacing $\Delta\lambda \sim N^{-2/3}$. The crossover from the KPZ regime to the ballistic regime occurs when the walker's resolution becomes comparable to this intrinsic gap, i.e., $1/\tau_p \sim \Delta\lambda$, which predicts $\tau_p^{\max} \sim N^{2/3}$. Our data for $N=64$ are consistent with this prediction, showing the onset of ballistic behavior for $\tau_p \gtrsim 64$.

\section{Simulation parameters}
\label{app:C}

\begin{table}[h]
\centering
\caption{Simulation parameters for each system size and temperature.}
\label{tab:params}
\begin{tabular}{c|c|c|c|c}
$N$ & $T$ & $E_0$ & $\text{base\_time}$ (steps) & $n_{\text{trajs}}$ \\ \hline
16 & 1.0 & -1.93 & 8000 & 50 \\
32 & 1.0 & -1.98 & 10000 & 45 \\
48 & 1.0 & -1.93 & 15000 & 50 \\
64 & 1.0 & -1.93 & 20000 & 50 \\
128& 1.0 & -1.93 & 30000 & 40 \\
64 & 0.5 & -1.83 & 20000 & 40 \\
64 & 1.5 & -1.93 & 15000 & 40 \\
64 & 2.0 & -1.93 & 15000 & 40 \\
\end{tabular}
\end{table}

The persistence times $\tau_p$ range from $1$ to $512$ in approximately logarithmic spacing (37 values). The time step is $\Delta t = 0.01$. For $\tau_p > 32$, the simulation time is extended as $T_{\mathrm{max}} = \text{base\_time} \times \sqrt{\tau_p/32}$, capped at $40000$ steps.

Fyodorov's static theory\cite{fyodorov2014, fyodorov2015} predicts that the $p=3$ energy landscape consists of a scale‑free hierarchy of high‑index saddles (index $\sim N/2$), offering abundant descent directions. To test this, we define the escape probability $P_{\mathrm{esc}}(\tau_p)$ as the fraction of trajectories that, starting from $E_0 = E_{\mathrm{th}}-0.3$ (see Table~\ref{tab:params}), reach the threshold $E_{\mathrm{th}} = -2\sqrt{2/3}$ within the simulation time.

For $p=3$, $P_{\mathrm{esc}}(\tau_p)$ shows no sharp resonance with $\tau_p$; values fluctuate around $0.25$–$0.35$ without systematic trend. This is consistent with a landscape where escape is largely independent of the walker's persistence time. In contrast, the $p=2$ landscape—a single narrow canyon—would yield a pronounced peak at an optimal $\tau_p^*$, a feature absent in $p=3$.

\section{Choice of fitting interval for KPZ exponent extraction}
\label{app:D}

The KPZ exponent $\alpha$ is extracted by fitting the energy correlation time $\tau_{\mathrm{corr}}$ as a function of the persistence time $\tau_p$ on a log-log scale. The fitting interval is chosen as $\tau_p \in [2,32]$. This choice is motivated by the following considerations.

\textit{Lower bound.} For $\tau_p = 1$, the dynamics exhibits systematic deviations from the asymptotic power law, as the correlation time is comparable to the simulation time step ($\Delta t = 0.01$), leading to larger statistical uncertainties. Moreover, this regime is close to the overdamped limit, where the scaling behavior is not yet fully developed.

\textit{Upper bound.} For $\tau_p > 32$, the system begins to cross over to the ballistic regime. The effective exponent $\alpha_{\mathrm{eff}} = d\log\tau_{\mathrm{corr}}/d\log\tau_p$ (computed from the data) starts to deviate from $1/3$ for $\tau_p \gtrsim 64$, indicating the onset of ballistic dynamics. Including these larger $\tau_p$ values would bias the fitted exponent toward higher values.

\textit{Stability and quality of fits.} Over the interval $\tau_p \in [2,32]$, the power-law fits consistently yield high $R^2$ values: $R^2 > 0.72$ for all $N$, and for the $N=64$ temperature scan, $R^2 > 0.98$ at $T=0.5$ and $T=2.0$. The extracted $\alpha$ values are stable under small variations of the fitting interval boundaries, confirming the robustness of the choice.

Therefore, the interval $\tau_p \in [2,32]$ is used throughout this work for extracting the KPZ exponent $\alpha$ in the $p=3$ spherical spin glass.

\end{document}